\documentclass[aps,prl,reprint,amsmath,amssymb]{revtex4-1}
\usepackage{graphicx}% Include figure files
\usepackage{dcolumn}% Align table columns on decimal point
\usepackage{bm}% bold math
\usepackage{color}

\usepackage[utf8]{inputenc}
\usepackage[T1]{fontenc}

\newcommand{\exciting}{{\usefont{T1}{lmtt}{b}{n}exciting}}
\usepackage{tcolorbox}
\definecolor{reddish}{rgb}{0.6,0,0}
\definecolor{blueish}{rgb}{0,0,0.5}
\definecolor{greenish}{rgb}{0,0.6,0}

\begin{document}

\draft % marks overfull lines with a black rule on the right

%\preprint{}

\title{Resta-like preconditioning for self-consistent field iterations in the linearized augmented planewave method} %Title of paper

\author{Jongmin Kim}
\email{jongmin.kim@uni-tuebingen.de}
%\thanks{*}
\affiliation{Institute of Theoretical Chemistry, Universität Ulm, Lise-Meitner-Straße 16, 89081 Ulm, Germany}
\affiliation{Institute of Physical and Theoretical Chemistry, Universität Tübingen, Auf der Morgenstelle 15, 72076 Tübingen, Germany}
\author{Matthias M.~May}
\affiliation{Institute of Theoretical Chemistry, Universität Ulm, Lise-Meitner-Straße 16, 89081 Ulm, Germany}
\affiliation{Institute of Physical and Theoretical Chemistry, Universität Tübingen, Auf der Morgenstelle 15, 72076 Tübingen, Germany}


\date{\today}

\begin{abstract}    
Convergence in self-consistent-field cycles can be a major computational bottleneck of density-functional theory calculations. We propose a Resta-like preconditioning method for full-potential all-electron calculations in the linearized augmented planewave (LAPW) method to smoothly converge to self-consistency. We implemented this preconditioner in the \exciting\ code and apply it to the two semiconducting systems of MoS$_2$ slabs and P-rich GaP(100) surfaces as well as the metallic system Au(111), containing a sufficiently large amount of vacuum. Our calculations demonstrate that the implemented scheme performs reliably as well as more efficiently regardless of system size, suppressing long-range charge sloshing.
{While the suitability of this preconditioning higher for semiconducting systems, the convergence for metals is only slightly decreased and thus still trustworthy to apply. Furthermore, a mixing algorithm with the preconditioner shows an improvement over that with the Kerker preconditioner for the investigated semiconducting systems.}
\end{abstract}

%\pacs{}% insert suggested PACS numbers in braces on next line

\maketitle 

\section{Introduction}
Kohn-Sham density-functional theory (DFT) has become one of the standard \textit{ab initio} approaches to predict electronic properties of materials \cite{Hohenberg1964, Kohn1965}. A major challenge in determining the ground-state charge density arises from the nonlinearity of the Kohn-Sham equations. This obstacle can be solved by considering a fixed-point iteration technique, also referred to as the self-consistent-field (SCF) iteration. During the SCF cycle, the Kohn-Sham potential, $v_\mathrm{KS}^\mathrm{in}$, is calculated using a given charge density $\rho^\mathrm{in}$ subsequently one can solve the Kohn-Sham equations. From this, a new density $\rho^\mathrm{out}$ is generated, serving as the input for the next SCF iteration. This iterative way is terminated when self-consistency to a given convergence threshold is achieved.

A tactical approach for efficient SCF convergence is mixing $\rho_{i}^\mathrm{in}$ and $\rho_{i}^\mathrm{out}$ of previous steps to construct a new density $\rho_{i+1}^\mathrm{in}$ in the next step. In the last decades, various mixing methods and their variations have been reported for electronic-structure theory \cite{Broyden1965, Johnson1988, Pulay1980, Pulay1982, Gonze1996, Vanderbilt1984, Ho1982, Marks2008, Bowler2000, Fang2009, Marks2013, Pratapa2015, Banerjee2016, Marks2021}, and they have proven to be successful in many systems. However, one often faces poor convergence -- and hence large computational costs -- due to fundamental challenges. A common problem is the charge sloshing instability that induces oscillations of the charge density during the SCF iteration \cite{Woods2019, Anglade2008, Kerker1981, Kresse19961, Kresse19962, Kohyama1996, Shiihara2008, Sundararaman2017, Zhou2018, Winkelmann2020, Kim2020}. This issue occurs when a unit cell becomes large, and it is generally more pronounced in metallic systems. One of the major sources of the charge sloshing is an element of $G^{-2}$ in the Hartree potential at long-wavelengths.

Different methods exist to overcome this shortcoming~\cite{Anglade2008, Kerker1981, Kresse19961, Kresse19962, Raczkowski2001, Vanderbilt1984, Ho1982, Auer1999, Sawamura2004, Lin2013, Herbst2021}. Among them, the Kerker preconditioner improves the convergence to alleviate the charge sloshing
\cite{Kerker1981}. This preconditioner is known that it is suitable for metallic systems. 
However, the question arises to what extent the preconditioner is applicable in the case of insulating or semiconducting, or inhomogeneous systems. References~\citenum{Herbst2021} and \citenum{Kumar2020} have reported that the Kerker preconditioner does not ensure the convergence for those systems because this preconditioner cannot properly describe their long-range screening~\cite{Zhou2018}. An alternative preconditioner to render the self-consistency procedure efficient for such systems is the Resta preconditioner~\cite{Resta1977}. In practice, both preconditioners are computationally less demanding compared to other methods for large unit cells. They were originally devised for planewave based methods. Recent theoretical studies have shown a formulation of the Kerker preconditioner as well as its implementation in full-potential (FP) calculations with the linearized augmented planewave (LAPW) basis set~\cite{Winkelmann2020, Kim2020}. The Resta preconditioner, however, is not yet reformulated to implement in the FP-LAPW method. 

In this paper, we develop a Resta-like preconditioning scheme to be applicable in the FP-LAPW method for stable and rapid SCF convergence. The Resta-like preconditioner is implemented in \exciting\ code \cite{Gulans2014}. We examine the performance of our Resta scheme and compare it with another preconditioner using examples of insulator and semiconducting systems: MoS$_2$ slabs and P-rich, $p(2\times2)$ GaP(100) surfaces. We find that the implemented preconditioner improves robustness and accelerates overall convergence of the self-consistency iterations, which would extend the range of systems accessible to FP-LAPW with a given computational power.

\section{Preconditioners: Kerker and Resta  \label{sec:kerkerrestea}}

The simplest mixing approach is to linearly mix the previous input and output charge density. In this case, a new guess at ($i$+1)th iteration, $\rho^\mathrm{in}_{i+1}$, is defined as 
\begin{equation}
\label{eq:linear}
\rho^\mathrm{in}_{i+1}(\mathbf{G}) = \rho^\mathrm{in}_i(\mathbf{G}) + \alpha \left[ \rho^\mathrm{out}_{i}({\bf G}) - \rho^\mathrm{in}_{i}({\bf G}) \right],
\end{equation}
where $\alpha$ is a damping parameter and appropriate values lie in the range between zero and one \cite{Baran2008}. $\alpha$ generally relies on the considered material, for instance, a small value of $\alpha $ is suitable for metallic systems \cite{Annett1995}. Using this simple method, SCF convergence can already be improved in some cases. However, in many structures, particularly metallic, large-scale, and inhomogeneous ones, poor convergence or even divergence is observed, even if one uses the optimal $\alpha$. In practice, strong fluctuations in $\rho_{i}^{\mathrm{out}}$ caused by a small change of $\rho_{i}^{\mathrm{in}}$ prevent the convergence during the self-consistency iteration. This phenomenon is well-known as charge sloshing \cite{Woods2019, Anglade2008, Kerker1981, Kresse19961, Kresse19962, Kohyama1996, Shiihara2008, Sundararaman2017, Zhou2018, Winkelmann2020, Kim2020}, which is more severe in metals with large supercells.  To avoid this problem, one can add an effective preconditioner in Eq.~\ref{eq:linear} in the following manner: 
\begin{equation}
\label{eq:precondition}
\rho^\mathrm{in}_{i+1}(\mathbf{G}) = \rho^\mathrm{in}_i(\mathbf{G}) + \alpha\, \mathrm{P}(\mathbf{G}) \left[ \rho^\mathrm{out}_{i}({\bf G}) - \rho^\mathrm{in}_{i}({\bf G}) \right],
\end{equation}
with $\mathrm{P}$ being the preconditioner. 

A number of preconditioners have been proposed to suppress the charge sloshing and to improve the SCF convergence \cite{Kerker1981, Anglade2008, Resta1977, Ho1982, Auer1999, Sawamura2004, Lin2013, Herbst2021}.  In this paper, we focus on two types of preconditioner: Kerker and Resta preconditioners \cite{Kerker1981, Resta1977}. Such preconditioners can be simply applied to common mixing schemes as multipliers and are computationally cheaper. 

The Kerker preconditioner is typically used to capture the long-range screening behavior, causing a stable performance of the self-consistency procedure, especially for metallic systems with large unit cells. This preconditioner has been developed on base of the Thomas-Fermi screening model, and it reads
\begin{equation}
\label{eq:kerkerpreconditioner}
\mathrm{P}({\bf {G}})_{\mathrm{Kerker}}=\frac{G^2}{G^2+\lambda^2},
\end{equation}
where $\lambda$ is a parameter that controls the screening at long wavelengths. For instance, $\mathrm{P}({\bf {G}})_{\mathrm{Kerker}}$ corresponds to one when $\lambda$ is zero. On the other hand, the value of $\mathrm{P}({\bf {G}})_{\mathrm{Kerker}}$ is prone to reduce with increasing $\lambda$ at short wavevectors $\mathrm{\bf{G}}$. {The Thomas-Fermi screening wave vector $k_{\mathrm{TF}}$ has been proposed as this parameter \cite{Zhou2018}, and it can be written by
\begin{equation}
\label{eq:thomas-fermi}
\lambda=k_{\mathrm{TF}}\sim \sqrt{4\pi N(\varepsilon_{\mathrm{F}})}.
\end{equation}
with $N(\varepsilon_{\mathrm{F}})$ being the density of states at the Fermi energy $\varepsilon_{\mathrm{F}}$.   
}

As discussed above, the ideal systems for the Kerker preconditioner are commonly metals, which can be treated in the picture of a homogeneous electron gas \cite{Zhou2018, Herbst2021}. Unfortunately, the use of such a preconditioner does not guarantee achieving smooth convergence for insulators or semiconductors, or inhomogeneous systems like surfaces \cite{Herbst2021, Kumar2020} as it does not capture properly the incomplete screening in those systems. Thus, one considers the Resta preconditioner, which can be used for those systems. The Resta preconditioner is given by 
\begin{equation}
\label{eq:resteapreconditioner}
\mathrm{P}({\bf {G}})_{\mathrm{Resta}}=\frac{\frac{{\kappa}^2 \mathrm{sin}(|{\bf G}| \beta)}{\gamma |{\bf G}|\beta}+{G}^2}{G^2+\kappa^2},
\end{equation}
where $\beta$, $\kappa$ and $\gamma$ are parameters. According to previous studies \cite{Zhou2018, Shajan1992, Mott1936}, those parameters are related to the screening length, Fermi-momentum-related quantity, and static dielectric constant, respectively.
As suggested by Resta~\cite{Resta1977}, one can obtain the static dielectric constant using other parameters as follows:
\begin{equation}
\label{eq:gamma}
\gamma=\frac{\mathrm{sinh}(\kappa\beta)}{\kappa\beta}.
\end{equation}
Here, $\kappa$ can be expressed as
\begin{equation}
\label{eq:kappa}
\kappa=\frac{(k_{\mathrm{F}}/\pi)^{0.5}}{2},
\end{equation}
where $k_{\mathrm{F}}$ is the valence electron Fermi momentum, and is defined as 
\begin{equation}
\label{eq:kf}
k_{\mathrm{F}}=(3\pi n_{0})^{1/3}
\end{equation}
with $n_{0}$ being the valence electron density. Finally, we estimate $\beta$ as the lattice constant of a system. {Since those parameters are already determined for the specific system, no further adjustment to choose their optimal values is required.}

In the Kerker and Resta preconditoner cases, it is straightforward and efficient to implement them in planewave-based methods, since they are initially devised in reciprocal space, but that is not straightforward for other methods like the FP-LAPW method. To implement these preconditioners in the FP-LAPW scheme, we firstly discuss the expression of the charge density. Here, the space of a unit cell consists of an interstitial region, \textit{I}, as well as atomic spheres around the centers of atoms, which are called muffin-tin spheres. The muffin-tin part of the charge density is expanded in terms of spherical harmonics, while the density is represented by planewaves in the interstitial part:     
\begin{equation}
\label{eq:LAPWdensity}
\rho({\bf r}) = 
\begin{cases}
\ \ \sum\limits_{ {\bf G}} \rho_{I}({\bf G}) \ e^{i{\bf G}{\bf r}} & {\bf r} {\in I}\\
\ \ \sum\limits_{lm} \rho^{\alpha}_{lm}({\bf r}) \ Y_{lm}({\bf \widehat{r}}) & {\bf r} {\in \mathrm{MT}_{\alpha}}.
\end{cases}
\end{equation}

To make such preconditioners compatible with the FP-LAPW method, we need to transform them into the real space representation. For this reason, we can rewrite Eqs.~\ref{eq:kerkerpreconditioner} and \ref{eq:resteapreconditioner} as
\begin{equation}
\label{eq:kerkerRS}
\mathrm{P}({\bf {r}})_{\mathrm{Kerker}}=1+\lambda^2\left(\nabla^2-\lambda^2 \right)^{-1},
\end{equation}
and 
\begin{equation}
\label{eq:restaRS}
\begin{aligned}
\mathrm{P}({\bf {r}})_{\mathrm{Resta}} \approx  &\, 1-\frac{\beta^2 \kappa^2}{6\,\gamma}\\
&+\left[\kappa^2- \frac{\kappa^2}{\gamma}-\frac{\beta^2 \kappa^4}{6\,\gamma}\right]\left(\nabla^2-\kappa^2 \right)^{-1},
\end{aligned}
\end{equation}
respectively. 
It should be noted that we approximate $\frac{{\kappa}^2 \mathrm{sin}(|{\bf G}| \beta)}{\gamma |{\bf G}|\beta}$ in Eq.~\ref{eq:resteapreconditioner} via a Taylor expansion, and this is truncated before the quadratic term. A further challenge that makes the proconditioner in Eqs.~\ref{eq:kerkerRS} and \ref{eq:restaRS} difficult to develop is how to apply the operator, $(\nabla^2-\lambda^{2} (\mathrm{or}\,\kappa^2) )^{-1}$, directly for the FP-LAPW case. Alternatively, we consider the screened Coulomb potential $V(\mathbf{r})$ that is the solution of the screened Poisson equation: 
\begin{equation}
\label{eq:screened}
\left(\nabla^2-\lambda^2 \right)V(\mathbf{r})=-4\pi\rho(\mathbf{r}).
\end{equation}
The input density $\rho(\mathbf{r})$ in Eq.~\ref{eq:screened} can be expressed it in terms of the residual obtained from a mixing approach, i.e., $\rho(\mathbf{r})=-\left[\rho^\mathrm{out}_{i}({\bf r}) - \rho^\mathrm{in}_{i}({\bf r})\right]/4 \pi$. Using them, $V(\mathbf{r})$ can be computed eventually.

We can solve Eq.~\ref{eq:screened} by employing Weinert’s pseudo-charge method \cite{Weinert1981}.
The method of the original work of Weinert is designed to obtain the solution of the Poisson equation in FP-LAPW calculations.
Based on this method, thus, the Hartree potential can be evaluated. 
Several studies have shown that it is still available for solving the screened Poisson equation \cite{Tran2011, Kim2020, Winkelmann2020}.
Tran \textit{et al}. used this method for the implementation of screened hybrid functionals \cite{Tran2011}. Recently, Kim \textit{et al}. and Winkelmann \textit{et al}. successfully implemented the Kerker preconditoner by utilizing the method \cite{Kim2020, Winkelmann2020} for FP-LAPW calculations. 

According to this method, a smooth charge density $\tilde{\rho}^\alpha(\mathbf{r})$ is used instead of the input charge density in muffin-tin spheres, therefore, the modified charge density can be expressed as 
\begin{equation}
\label{eq:pseudodensity}
\bar{\rho}(\mathbf{r})=\sum\limits_{\mathbf{G}}\rho_\mathbf{G} \ e^{i\mathbf{G}\mathbf{r}}+\sum\limits_{\alpha} \tilde{\rho}^\alpha(\mathbf{r}),
\end{equation}
where the first term of the right-hand side is the density in the interstitial region. Equation~\ref{eq:pseudodensity} then allows to perform a Fourier transformation. Now, one can calculate the screened potential in the interstitial region:

\begin{equation}
%\label{eq:screenedG}
V_{I}({\bf G})=  \frac{4\pi}{G^2+\lambda^2}\,\,\bar{\rho}({\bf G}).
\end{equation}
The Fourier component of $\bar{\rho}({\bf r})$ is analytically defined as 
\begin{equation}
\begin{aligned}
\tilde{\rho}^{\alpha}({\bf G}) = &\frac{4\pi}{\Omega} e^{-i{\bf G}\cdot{\bf R}^{\alpha}} \sum\limits_{l=0}^\infty\sum\limits_{m=-l}^l \frac{(-i)^{l}}{(2l+1)!!} \\ 
&\times \frac{\lambda^{l+n+1} j_{l+n+1} (G R_{\alpha}) }{i_{l+n+1} (\lambda R_{\alpha}){G}^{n+1}}Y_{lm}({\bf \widehat{G}}) \tilde{q}_{lm}^{\alpha},
\end{aligned}
\end{equation}
with $j_{l}$ and $i_{l}$ being the spherical Bessel functions and modified spherical Bessel functions of the first kind. We, in turn, determine the potential inside muffin-tin spheres by solving the Dirichlet boundary-value problem, and it reads 
\begin{equation}
\label{eq:mtsolution}
V^{\alpha}({\bf r}) = \int_{MT_\alpha} G({\bf r}, {\bf r'})  \rho({\bf r'}) d{\bf r'} - \frac{R_{\alpha}^2}{4\pi} \oint_{S_\alpha}V_{I}(R_{\alpha})  \frac{\partial G}{\partial n'} d\Omega'.
\end{equation}
Here, $G(\bf{r})$ is a Green-function that is given as
\begin{equation}
\begin{aligned}
G({\bf r}, {\bf r'}) = 4\pi\lambda\sum\limits_{l=0}^\infty\sum\limits_{m=-l}^l i_{l}(\lambda r_{<})k_{l}(\lambda r_{>}) \\
\times \left[1-\frac{i_{l}(\lambda r_{>})k_{l}(\lambda R_{\alpha})}{k_{l}(\lambda r_{>})i_{l}(\lambda R_{\alpha})}\right]Y_{lm}^{*}({\bf \widehat{r}'})Y_{lm}({\bf \widehat{r}}),
\end{aligned}
\end{equation}
where $k_{l}$ denote the modified spherical Bessel functions of the second kind. Detailed derivations of the pseudo-charge method for the screened Poisson equation are provided in Refs.~\citenum{Kim2020}, \citenum{Tran2011}, and \citenum{Winkelmann2021}.

\section{Pulay mixing \label{sec:pulay}}
In general, Pulay mixing \cite{Pulay1980, Pulay1982}, which is called \textit{direct inversion in the iterative subspace} (DIIS), performs better than linear mixing. The main difference between Pulay and linear mixings is that an iterative history of the input densities $\rho^\mathrm{in}_i(\mathbf{r})$ and residuals $R^\mathrm{in}_i(\mathbf{r})$, which are the difference between $\rho^\mathrm{out}_i(\mathbf{r})$ and $\rho^\mathrm{in}_i(\mathbf{r})$, is stored for the former scheme, determining optimum densities and residuals: 
\begin{equation}
\label{eq:optimal}
\begin{aligned}
R_{\mathrm{opt}}^{\mathrm{in}}=\Sigma_i\omega_i R_i^{\mathrm{in}}\\
\rho_{\mathrm{opt}}^{\mathrm{in}}=\Sigma_i\omega_i \rho_i^{\mathrm{in}}.
\end{aligned}
\end{equation}
Minimizing the norm of the residual is required for the optimum residual.
To do so, weights $\omega_{i}$ in Eq.~\ref{eq:optimal} satisfy the constraint as 
\begin{equation}
\label{eq:constraint}
\Sigma_i\omega_i=1.
\end{equation}
These weights can be obtained by solving a system of linear equations \cite{Woods2019}:
\begin{equation}
\begin{pmatrix}
R^{\dagger}_{1}R_{1} & R^{\dagger}_{1}R_{2} & \cdots & R^{\dagger}_{1}R_{m} & 1 \\
R^{\dagger}_{2}R_{1} & R^{\dagger}_{2}R_{2} & \cdots & R^{\dagger}_{2}R_{m} & 1 \\
\vdots & \vdots & \vdots & \ddots & \vdots \\
R^{\dagger}_{m}R_{1} & R^{\dagger}_{m}R_{2} & \cdots & R^{\dagger}_{m}R_{m} & 1 \\
1 & 1 & \cdots & 1 & 0
\end{pmatrix}
\begin{pmatrix}
\omega_1 \\
\omega_2 \\
\vdots \\
\omega_{m} \\
\lambda
\end{pmatrix}
=
\begin{pmatrix}
0 \\
0 \\
\vdots \\
0 \\
1
\end{pmatrix}
,
\end{equation}
with $\lambda$ being a Lagrange multiplier.
With a preconditioner $\mathrm{P}$, the new input density for the next iteration in the Pulay mixing is updated as follows:
\begin{equation}
\label{eq:Pulay}
\rho_{i+1}^{\mathrm{in}}= \rho_{\mathrm{opt}}^{\mathrm{in}}+\alpha \mathrm{P} R_{\mathrm{opt}}^{\mathrm{in}},
\end{equation}

where $\alpha$ is the same parameter discussed for the linear mixing case.
In this work, we also consider Kerker and Resta-like preconditioners for this mixing.
It should be noted that one can exploit the same approach of the pseudo-charge method explained in  Sec.~\ref{sec:pulay} when these preconditioners are involved in the second term of the right-hand side of Eq.~\ref{eq:Pulay}.

\section{Computational details}
\begin{figure}
\begin{center}
\includegraphics[width= .8\linewidth]{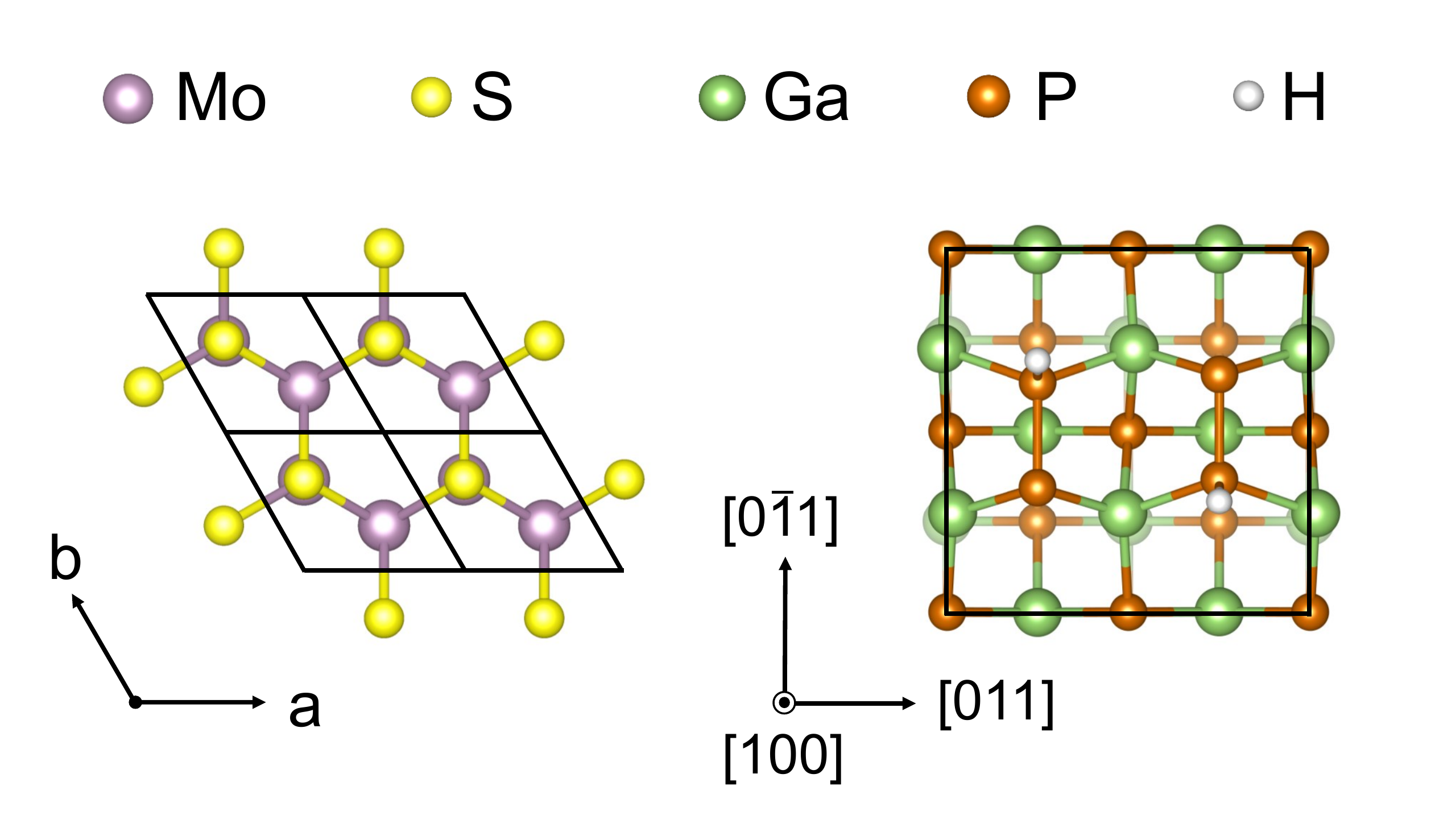}
\end{center}
\caption{Left: Top view of a MoS$_2$ slab (left) and a P-rich, $p(2\times2)$ GaP(100) surface. The unit cells are represented by the black lines.} 
\label{fig:structure}
\end{figure} 
As test cases, the two systems of MoS$_2$ slabs and P-rich, $p(2\times2)$ GaP(100) surfaces are considered.  We depict both structures in Fig~\ref{fig:structure}.
In particular, we model 10 layers (30 atoms) and 20 layers (60 atoms) for the former, which has an AB stacking pattern, and 15 layers (64 atoms) and 19 layers (80 atoms) for the latter. In the latter structures, a surface reconstruction is formed, containing buckled P--P dimers on top, together with one hydrogen atom per dimer. 
{For the sake of comparison, we also consider the Au(111) surface (with 15 layers) as generic example for a metallic system.}
Such structures are constructed with lattice constants of 3.19\,\AA{}, 5.45\,\AA{}, and 4.19\,\AA{} for the MoS$_2$, GaP(100), and Au(111), respectively. As mentioned before, an elongated unit cell tends to exhibit slow SCF convergence due to the charge sloshing. We therefore set around 30\,\AA{} of vacuum along the perpendicular direction that also prevents spurious interactions between neighboring replica for both systems.

The aforementioned methods in Secs.~\ref{sec:kerkerrestea} and \ref{sec:pulay} are implemented in the full-potential all-electron density-functional theory (DFT) code \exciting. All calculations are performed using this code.  
We employ the muffin-tin radii $R_{MT}$ of 2.4, 2.1, 1.7, 1.5, 0.9, 1.9 bohr for Mo, S, Ga, P, H, and Au atoms, respectively.    
For the exchange-correlation functional, we adopt the generalized gradient approximation in the Perdew-Burke-Ernzerhof (PBE) parametrization \cite{Perdew1996}.
A basis-cutoff of $R_{MT}G_{max}=$~6.5, 3, and 7 are used for our MoS$_2$, GaP(100), and Au(111) cases.
Moreover, the sampling of the Brillouin zone (BZ) are carried out with a 21~$\times$~21~$\times$~1, a 6~$\times$~6~$\times$~1, and a 16~$\times$~16~$\times$~1  grid for the MoS$_2$ slabs, reconstructed GaP(100), and Au(111) surfaces.  
We assume that calculations are converged when the root mean squared (RMS) change of residual matrices is smaller than 10$^{-6}$ e/bohr$^{1.5}$ between two consecutive iterations.
The RMS is defined as
\begin{equation}
\mathrm{RMS}=\sqrt{\frac{\left||R^{\mathrm{in}}|\right|^{2}}{\Omega}},
\end{equation}
where $\Omega$ denotes the unit cell volume. 

In our investigated systems, the mixing parameter $\alpha$ is set to 0.4, and we fix a history length of densities as well as residuals obtained from previous steps to 15 for the case of Pulay mixing ($m=15$).
{We set $\lambda=0.529$~bohr$^{-1}$ of the Kerker preconditioner for the MoS$_2$ and GaP(100) systems. 
In the case of Au(111), Eq.~\ref{eq:thomas-fermi} is used for the determination of $\lambda$.}

\section{Results and discussion}
To verify the efficiency of our implemented methods, we first start with 10 layers of MoS$_2$ slab.
In this case, we use linear mixing and the corresponding mixings that include Kerker and Resta-like preconditioners (termed \textit{Kerker mixing} and \textit{Resta mixing}, respectively).
\begin{figure}
\begin{center}
\includegraphics[width=.8\linewidth]{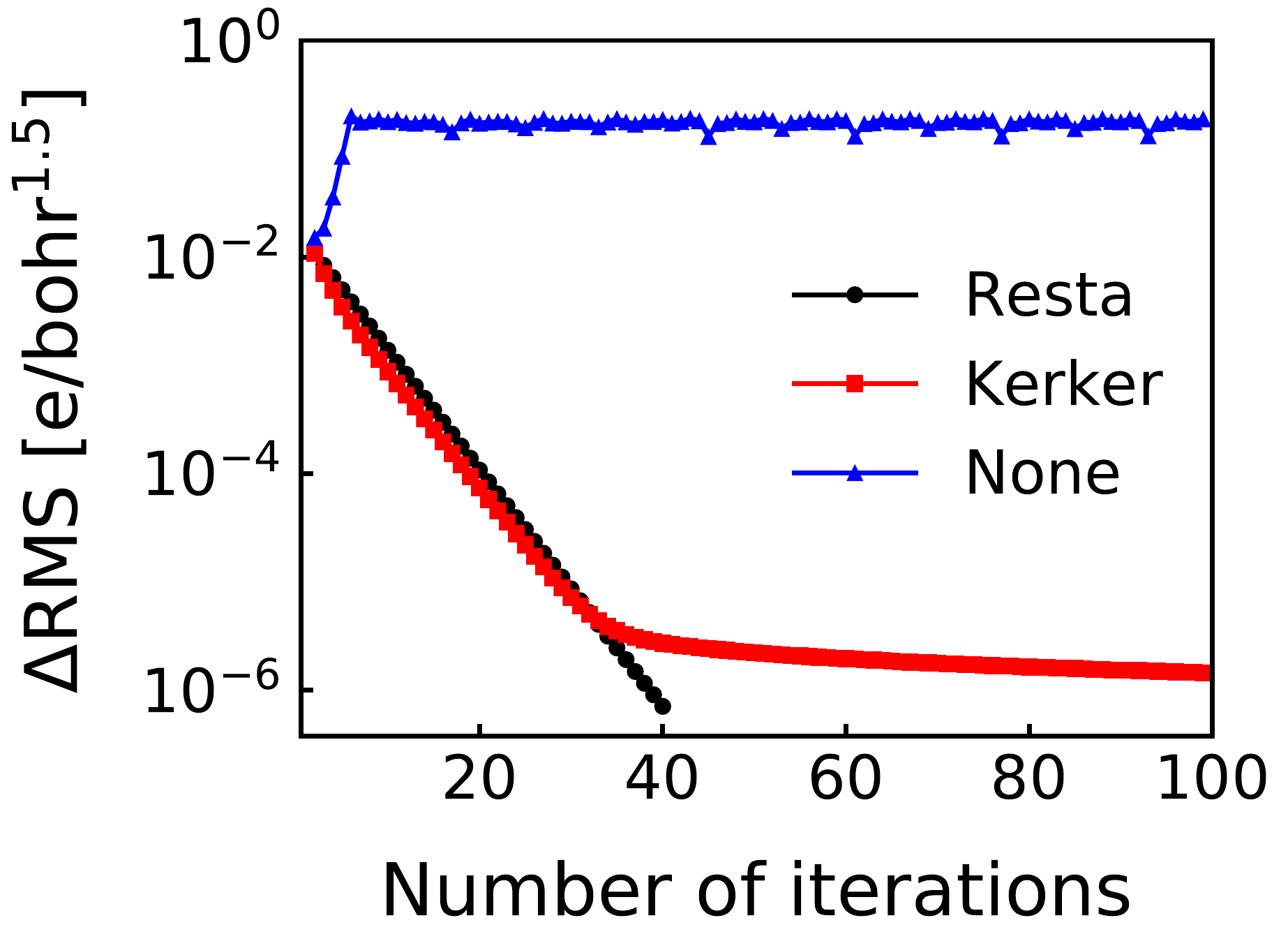}
\end{center}
\caption{The convergence of RMS change for 10 layers MoS$_2$ slab using linear mixing without and with Kerker and Resta-like precontitoners.} 
\label{fig:linkerkerresta}
\end{figure}

Figure~\ref{fig:linkerkerresta} exhibits the convergence behavior of the SCF iteration in terms of RMS change of residuals between two consecutive steps, $\Delta$RMS.
We find that the performance is highly sensitive to the type of preconditioner in this structure. It is clear that standard linear mixing (no preconditioner) does not reach the target threshold within 100 iterations. This scheme shows that the RMS fluctuation remains mainly around 10$^{-1}$ e/bohr$^{1.5}$. 
On the contrary, its performance is improved by employing preconditioners.
Both preconitioners show smoothly varying lines as shown in Fig.~\ref{fig:linkerkerresta}.    
The Kerker mixing, however, does not achieve the self-consistency during the first 100 steps.
This mixing decays $\Delta$RMS well until 35 steps, but it remains almost unchanged after that. Compared to other two methods, the Resta mixing converges to the target precision in 40 steps. Unfortunately, convergence is rather too slow to be practical. To further improve the efficiency, the Pulay mixing method with preconditioners is required.    
\begin{figure}
\begin{center}
\includegraphics[width=.8\linewidth]{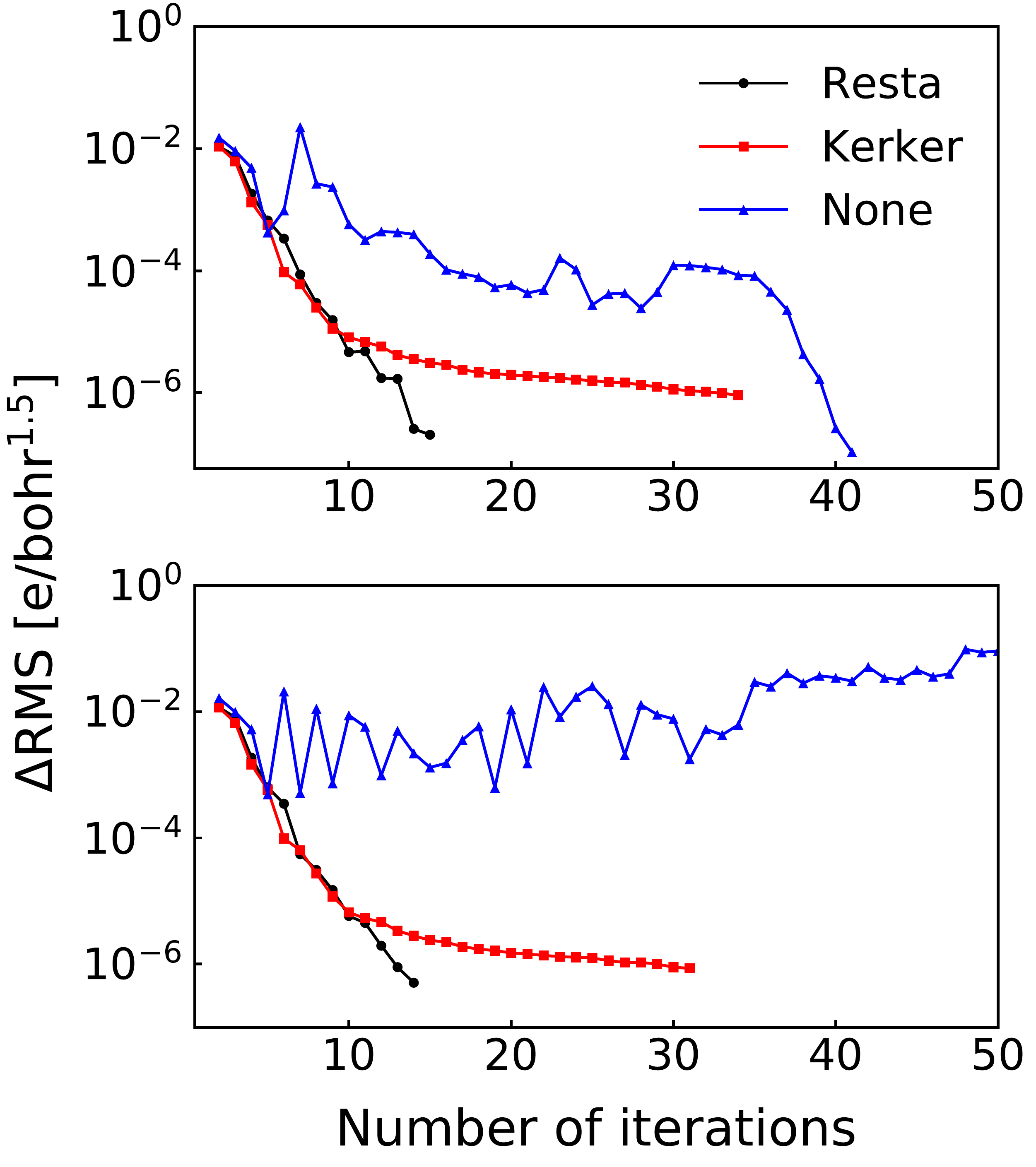}
\end{center}
\caption{The convergence of RMS change for 10 (top) and 20 (bottom) layers of MoS$_2$ slabs with standard Pulay and Pulay with Kerker and Resta-like preconditioners mixings.} 
\label{fig:pulaymos2}
\end{figure}

We observe that the slow convergence issue is tackled by Pulay-based approaches. This is displayed in the top panel of Fig.~\ref{fig:pulaymos2}. Like for our previous results, computed by linear, Kerker, and Resta mixing methods (see Fig.~\ref{fig:linkerkerresta}), we compare the convergence of $\Delta$RMS for three different cases, i.e., the Pulay without and with Kerker and Resta-like preconditioners. In contrast to linear and Kerker mixings, those three approaches achieve the convergence within 50 steps. The Pulay approach with the Resta-like preconditioner outperforms the other methods. This method converges in only 15 steps, and it is faster by at least 50~\% than the standard Pulay (no preconditioner), but also faster than with the Kerker preconditioner. 

To analyze the dependence of system size on the SCF convergence, we increase the number of layers to 20. The RMS convergence behavior is depicted in the bottom panel of Fig.~\ref{fig:pulaymos2}. Compared to the 10 layers case, the behavior of the Pulay mixing without any preconditioner is significantly changed, showing no convergence. This is attributed to the further elongated unit cell that causes the above-mentioned charge sloshing. Interestingly, the Pulay mixing methods with both preconditioners are insensitive to the system size. For instance, the required number of steps for the 20 layers case to reach the target precision is 14 in the Resta-like preconditioning approach, which is only one step less than in the case of 10 layers MoS$_2$. For the Kerker one, the difference of the convergence steps between both structures is three. 
This implies that the long-wavelength instability is magnificently suppressed. Overall, the Pulay with the Resta-like preconditioner is the most effective and efficient among discussed methods for our insulating system. 
\begin{figure}
\begin{center}
\includegraphics[width=.8\linewidth]{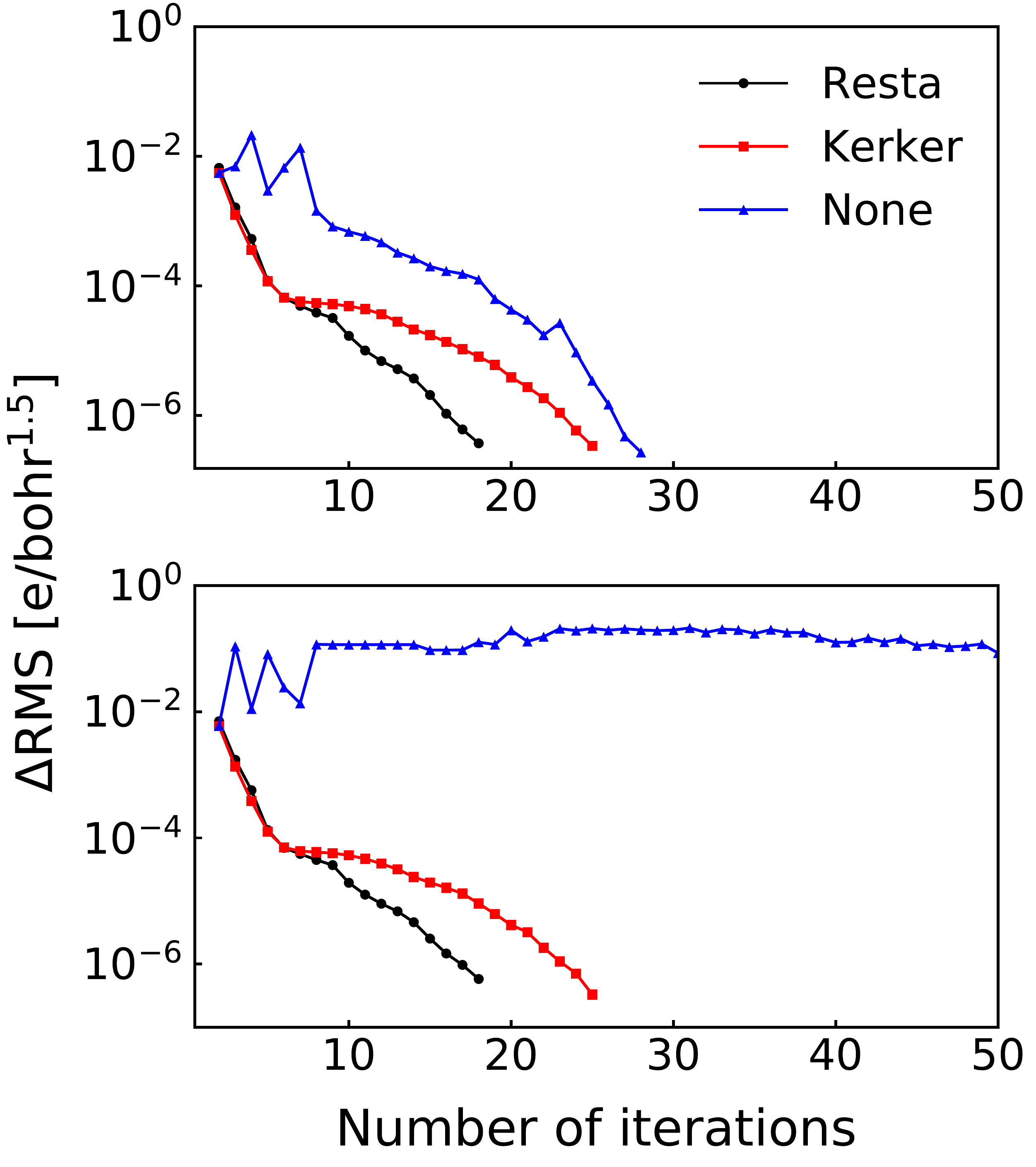}
\end{center}
\caption{The convergence of RMS change for 15 (top) and 19 (bottom) layers of P-rich, $p(2\times2)$-reconstructed GaP(100) surfaces  with standard Pulay and Pulay with Kerker and Resta-like preconditioners mixings.} 
\label{fig:pulaygap}
\end{figure}

Next, we choose P-rich, $p(2\times2)$ GaP(100) surfaces consisting of 15 and 19 layers as the second benchmark system with a classical surface reconstruction. In Fig.~\ref{fig:pulaygap}, for both structures, we present the RMS convergence using different methods, i.e., the Pulay without and with Kerker and Resta-like preconditioners during the self-consistency cycle. 
%Since reference~\citenum{Shajan1992} suggested relevant parameter values for the Resta preconditioner for bulk GaP, we adopt their suggested values: $\beta=~7.35$\,bohr, $\kappa=~0.58$\,bohr$^{-1}$, and $\gamma=~8.5$.
%Mixing methods discussed here, require overall smaller convergence steps because the charge sloshing is less pronounced to these systems, compared to our MoS2 systems.
Note that our results exhibit a similar tendency compared to those for the MoS$_2$ slabs. The conventional Pulay mixing has a dependence on the size of the system, while others are independent of it. We demonstrate that Pulay mixings with both preconditioner are practical since they converge in 15--30 steps, almost irrespective of the number of layers. The Pulay mixing combined with the Resta preconditioner is indeed superior to that with the Kerker one, saving around 30\,\% of the number of iterations. This indicates that such a scheme can capture the incomplete screening feature well in these surfaces. We conclude that the Resta-like preconditioner is more appropriate for use with insulators or semiconductors with large unit cells. 

\begin{figure}
\begin{center}
\includegraphics[width=.8\linewidth]{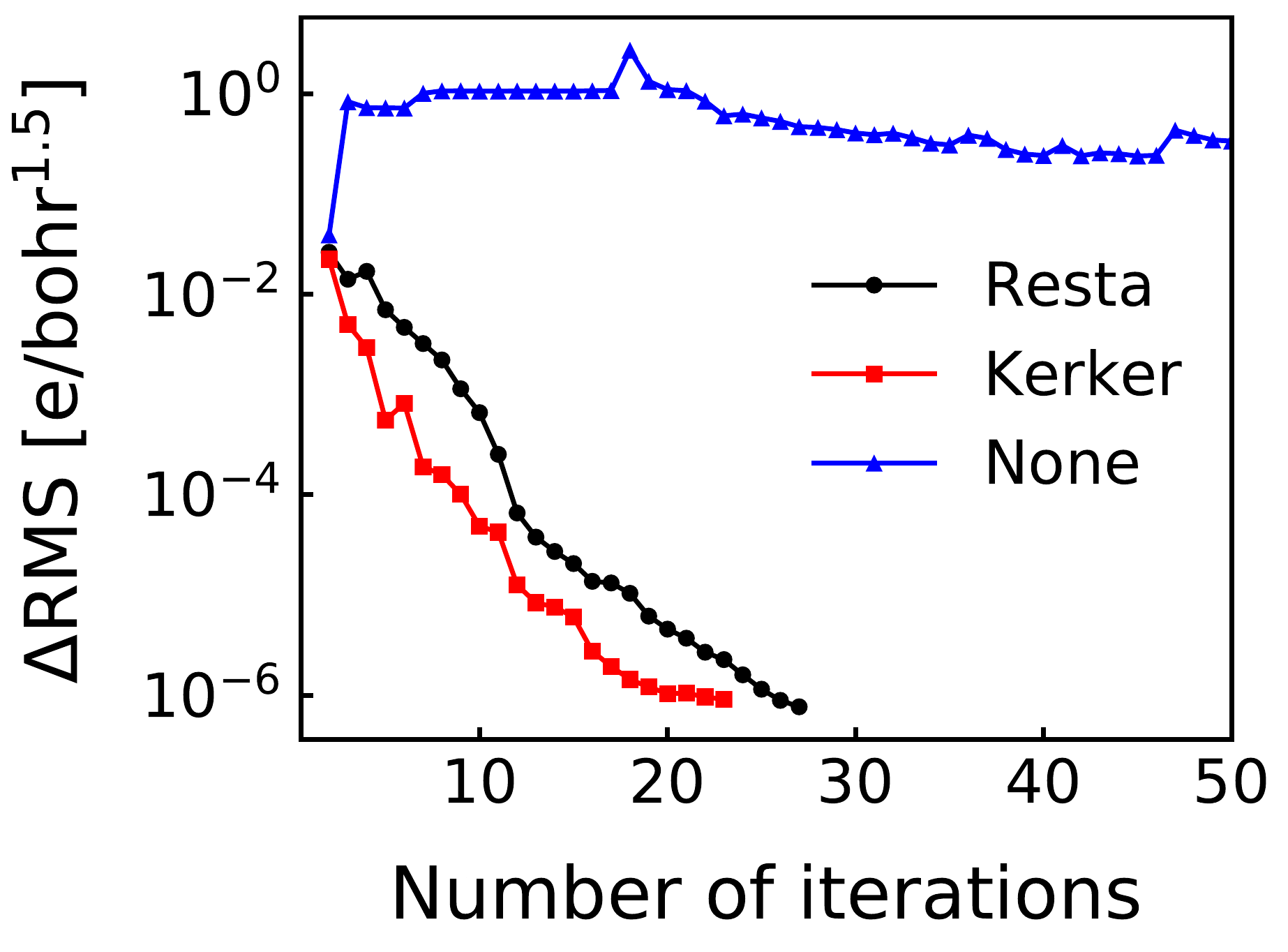}
\end{center}
\caption{{The convergence of RMS change for the Au(111) surface with standard Pulay and Pulay with Kerker and Resta-like preconditioners mixings.}} 
\label{fig:pulayau111}
\end{figure}

{In Fig.~\ref{fig:pulayau111}, we analyze the results of the convergence for the Au(111) surface. Similar to the MoS$_2$ and GaP(100) cases, the Pulay mixings with Kerker and Resta-like preconditioners exhibit robust convergence. On the other hand, unlike our previous calculations, the Kerker case is slightly faster than the Resta-like one. The difference of their steps is 4, corresponding to 15~\% more steps until convergence is reached. This is not surprising because the Resta model has been originally designed to describe screening in semiconductors. In principle, the static dielectric constant of metals is ideally infinite, yet our calculation yields a finite value that leads to deteriorating the performance. 
Despite the slightly slower convergence compared to Kerker, the Resta-like preconditioner is also applicable to metallic systems with a moderately slower convergence.}

\section{Summary and conclusions}
In this work, we have established a formulation of the Resta-like preconditioner that can be directly applied in the FP-LAPW method, and evaluated its performance for insulating and semiconducting systems. This algorithm has been implemented in the all-electron full-potential code \exciting. We have demonstrated that using this preconditioner in mixing methods leads to a more stable and faster SCF convergence than without preconditioner or with the Kerker preconditioner, and which is not sensitive to the system size. {The performance improvement for insulating and semiconducting systems is significant, while the performance loss for metallic systems is only moderate.} The increased robustness and efficiency extend the range of systems accessible in the FP-LAPW method, especially for materials of semiconducting or insulating nature as well as inhomomgeneous systems. 

\begin{acknowledgments}
We appreciate funding from the German Research foundation (DFG), No.~434023472.
The state of Baden-Württemberg, through bwHPC, and the German Research Foundation (DFG), through grant no.~INST 40/575-1 FUGG (JUSTUS 2 cluster), are acknowledged for providing the HPC resources.
\end{acknowledgments}

%\bibliography{ref}
%\section*{References}

\end{document}